\documentclass[prb,amsmath,twocolumn,floatfix,showpacs]{revtex4}
\usepackage{graphicx} \usepackage{latexsym}

\begin{document}

\title{ Rate constants for diffusive processes by partial
path sampling} \author{Daniele Moroni, Peter G. Bolhuis}
\affiliation{Department of Chemical Engineering, Universiteit van
Amsterdam, Nieuwe Achtergracht 166, 1018 WV Amsterdam, The
Netherlands} \author{Titus S. van Erp} \affiliation{Laboratoire de
Physique / Centre Europ\'een de Calcul Atomique et Mol\'eculaire,
Ecole Normale Sup\'erieure de Lyon, 46 all\'ee d'Italie, 69364 Lyon
Cedex 07, France}

\date{\today}

\begin{abstract}
We introduce a path sampling method for the computation of rate
constants for systems with a highly diffusive character. Based on the
recently developed algorithm of transition interface sampling (TIS)
this procedure increases the efficiency by sampling only parts of
complete transition trajectories confined within a certain region. The
algorithm assumes the loss of memory for highly diffusive progression
along the reaction coordinate.  We compare the new technique to the
TIS method for a simple diatomic system and  show that the
computation time of
the new method scales linearly, instead of quadraticaly, with the length of the diffusive barrier. The validity of the memory loss assumption is also
discussed.
\end{abstract}


\maketitle

\section{introduction}

The calculation of rate constants in complex systems by
straightforward molecular dynamics (MD) simulation is prohibited by the
exponential dependence of
the rate on the activation barrier height. The
expectation time for a reaction can easily be on the order of
milliseconds to seconds, whereas most simulation algorithms are
limited to molecular time-steps of femtoseconds. A single occurrence
of a rare event in a complex system can thus easily exceed current
computer capabilities by orders of magnitude.  The standard
Bennett-Chandler reactive flux method is able to avoid this timescale
problem by calculating the probability to be at the top of the
activation free energy barrier in combination with a time dependent
transmission coefficient~\cite{DC78,FrenkelSmit}. Although in
principle correct, the accuracy of this method is very sensitive to
the choice of reaction coordinate. In a complex reaction, an intuitive
simple reaction coordinate can lead to extremely low transmission
coefficients and, hence, an inaccurate or even immeasurable rate
constant.  Improving the reaction coordinate, by, for instance,
incorporating solvent degrees of freedom
(see e.g. Ref.~[\onlinecite{Timoneda91,Sprik2000}]),
is usually a very difficult task and requires a lot of a-priori
knowledge of the system.

The transition path sampling (TPS) technique of Chandler and
co-workers~\cite{TPS98,TPS98a,TPS99,Bolhuis02,Dellago02} does not need
any prior knowledge of the reaction coordinate and harvests a
collection of transition paths that connect the reactant with the
product states.  This ensemble of true dynamical paths allows detailed
understanding of the kinetics and mechanism of the reaction.  In
addition, the rate constant can be computed.  Processes as diverse as
cluster isomerization, auto dissociation of water, ion pair
dissociation, the folding of a polypeptide\cite{bolhuisPNAS} and
reactions in aqueous solution have been studied with TPS (see
Ref.~[\onlinecite{Bolhuis02}] for an overview).

As the TPS rate constant calculation is rather computer time
consuming, we recently introduced the transition interface
sampling (TIS) technique \cite{ErpMoBol2003}, thereby improving the
efficiency of the rate constant calculation substantially.  By
allowing a variable path length, TIS drastically reduces the number of
required time-steps for each path.  In addition, TIS is less sensitive
to recrossings and has a better convergence compare to the TPS rate constant method as it only counts
the effective positive terms.

In this paper, we will focus on transitions with a highly diffusive
character, or in the regime of high solvent friction.  
Examples are the
folding and unfolding of a protein in water, charge transfer,
fragmentation reactions, 
diffusion of a molecule through a membrane, 
and nucleation processes.  These types of processes have to overcome 
a relatively flat and wide, but still rough free energy barrier.
When applying the TPS (or TIS)  shooting algorithm to such a transition, the  Lyapunov instability causes the paths to diverge before the basins of attraction have the chance to  guide the paths to the proper stable state. Pathways  will then   become very long and, moreover, the acceptance ratio of shooting will be low. Hence, the shooting algorithm  will be very inefficient, resulting in bad sampling. 
Recently, we showed how to sample
long paths efficiently on a diffuse barrier with TPS by introducing
a little stochasticity in the trajectories~\cite{bolhuis03}.

Here, we will introduce an efficient method to calculate the rate constant for
such barriers.  To do so, we make use of the TIS effective flux
relation~\cite{ErpMoBol2003} and assume that the diffusivity
eliminates any memory effects over a distance more than the separation
between two interfaces.  The rate constant can then be recast in a
recursive relation for the hopping transition rates between
interfaces. These hopping transition rates can
be computed by sampling short trajectories connecting just three
successive interfaces.  If the assumption of memory loss is valid,
this partial path transition interface sampling (PPTIS) procedure
correctly collects the contributions of all possible paths to the rate
constant, in principle, even those with infinite lengths.

This paper is organized as follows:
In Sec.~\ref{sectheory}A we
illustrate the PPTIS concept for a simple one dimensional array of
well defined metastable states.  Although not completely without
physical importance, the model in this section can
only describe a
limited number of physical systems since most diffusive systems do not
have well defined metastable regions in the free energy landscape
between reactant and product state. In Sec.~\ref{TIStheory} we present
the TIS technique 
in a way that facilitates the
derivation of the rate expression for general diffusive barriers given
in Sec.~\ref{PPTIStheory}.  The implementation of the sampling
algorithm and the analysis of the accuracy of the assumptions are
discussed in Sec.~\ref{secalgorithm} and~\ref{secposint},
respectively.
A comparison between the TIS and PPTIS methods is made in Sec.~\ref{comparison}.
In Sec.~\ref{secNum}, we test the method and compare with the TIS
results for a isomerization reaction of a diatomic molecule with an
intrinsic long and flat barrier immersed in a fluid of repulsive
particles.  We end with concluding remarks and prospectives in
Sec.~\ref{secConclusions}.

\section{Theory}
\label{sectheory}

\subsection{Illustration of the PPTIS concept}
\label{illusex}

\begin{figure}[t]
\includegraphics[angle=-90,width=8cm]{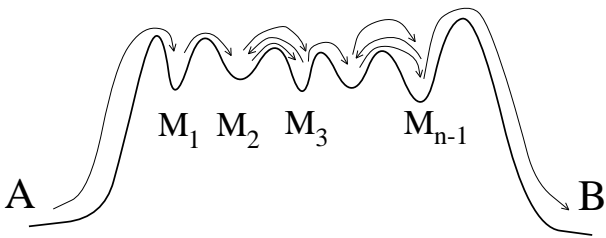}
\caption{Illustration of a barrier consisting of a series of
metastable states.  One possible trajectory connecting $A$ and $B$ is
shown.}
\label{metaseries}
\end{figure}

Before embarking on the general case of diffusive barriers, we will
first consider a simple one dimensional system that serves as an
illustrative example.  This system exhibits a barrier consisting of a
series of metastable states as is illustrated in
Fig.~\ref{metaseries}.
The overall barrier is high compared to those between metastable
states, so that the system shows two state kinetics and an overall
rate constant $k_{AB}$ is well defined.  We assume that the system can
hop from one metastable state to a neighboring one after which it will
fully relax.  Consequently, the probability to hop to left or right
does not depend on the history of the path, and hence the system
behaves Markovian. For this type of system, we might write down a
master equation and solve for all the population densities in each
state on the barrier as a function of time~\cite{vankampen}. However,
if we assume steady state behavior, and take into account the fact that the
population on the barrier is low, the overall rate constant is only
determined by the hopping probabilities.  We will denote the
probabilities to transfer from site $i$ to the right or left
metastable state by $t_{i,i+1}$ and $t_{i,i-1}$, respectively, which
are related by $t_{i,i+1}+t_{i,i-1}=1$.  For a system with $n-1$
metastable states: $M_1,M_2, \ldots M_{n-1}$ and the stable states
$M_0=A$ and $M_n=B$, the reaction rate $k_{AB}$ and its reverse
$k_{BA}$ can be expressed as:
\begin{eqnarray}
k_{AB}&=&k_{0,1} T[1 \rightarrow_0^n], \nonumber \\
k_{BA}&=&k_{n,{n-1}} T[n-1 \rightarrow_n^0],
\label{ratemeta}
\end{eqnarray}
with $T[i \rightarrow_m^j]$ the probability to reach via an arbitrary
number of hops from metastable state $i$ to metastable state $j$
before visiting metastable state $m$.  The computation of the rate
constants only requires the determination of the nearest neighbor
hopping probabilities $t_{i,i+1}$ and the first hopping rates
$k_{0,1}$ and $k_{n,{n-1}}$. The long distance hopping probabilities
$\{ T[1 \rightarrow_0^j], T[j-1 \rightarrow_j^0] \}$ can be obtained
via following recursive relations (see Appendix \ref{AppendixA}):

\begin{eqnarray}
T[1 \rightarrow_0^j] &=& \frac{ t_{j-1,j} T[1 \rightarrow_0^{j-1}] }{
t_{j-1,j}+t_{j-1,j-2}T[j-2 \rightarrow_{j-1}^{0}] }
\label{metahop} \\
T[j-1 \rightarrow_{j}^0] &=& \frac{t_{j-1,j-2}
T[j-2\rightarrow_{j-1}^0] }{ t_{j-1,j} + t_{j-1,j-2}
T[j-2\rightarrow_{j-1}^0] } \nonumber
\label{eq:recurs2}
\end{eqnarray}

Starting with $T[1\rightarrow_0^1]=T[0 \rightarrow_1^0]=1$, we can
iteratively solve 
Eqs.~(\ref{metahop}) for $j=2, 3 \ldots$ to $j=n$.
In this way we collect analytically the statistics of all possible
pathways. This procedure accounts for the straight-forward barrier
crossings, but also accounts for the contributions to the rate of an
infinite number of different pathways that approach from $A$ to $B$ in
an infinite number of hops.
Although the probability of a single
pathway is decreasing with its length, the total contribution of the
very long pathways becomes more important when $n$ is increased. In
fact, the average path length increases as $\sim n^2$.  In case of
uniform symmetric hopping ($t_{i,i+1}=t_{i,i-1}=\frac{1}{2}$ for all
$i$) one can show by induction that $k_{AB}=\frac{1}{n}k_{0,1}$,
whereas if we would only account the fastest pathway ($M_0 \rightarrow
M_1 \rightarrow M_2 \ldots \rightarrow M_n$) it would be much lower,
$(\frac{1}{2})^n k_{0,1}$.

At first sight, it seems a bit surprising that the residence time in
each metastable state and the absolute intra-barrier rates $k_{i,{i\pm
1}}$ have no influence on the final total rate expression.  Only the
relative rates are important as they determine the nearest neighbor
hopping probabilities by $t_{i,i\pm1}=k_{i,{i\pm 1}}/(k_{i,{i+
1}}+k_{i,{i- 1}})$.
Of course, when we start with a system out of equilibrium and
calculate the relaxation time from $A$ to $B$ for a system that is
initially completely in $A$, the intra-barrier rates $k_{i,{i\pm 1}}$
will be dominant factors.

Our treatment of this model can be related to the solution of Kramer's
equation if one considers a flat high barrier of length $l$.
Kramer's
equation then gives for the rate constant $k_{AB} \approx (D/l)
\exp(-\beta U)$ where $U$ is the barrier height and $D$ the diffusion
constant on top of the barrier\cite{FrenkelSmit}.  The connection
becomes clear when one realizes
$k_{0,1}/k_{1,0} = \exp(-\beta U)$ and $D/l = k/n$, with $k\sim
k_{1,0}$ the hopping rate, and $n$ the number of hops on the barrier.
Hence,
$k_{AB} = \frac{1}{n} k_{0,1}$, just as found above for the symmetric
uniform hopping model. A more formal treatment of general diffusive
Markov processes can be found in e.g. Ref.~[\onlinecite{vankampen}].

The model described above is of limited importance due to its highly
symmetric and one-dimensional character.  Some processes, however,
such as the diffusion of particles through a one-dimensional crystal
(e.g alkanes through zeolites) can be described by this uniform
symmetric hopping model.  More complex behavior such as diffusion
on surfaces, 
through multi-dimensional crystals, or in (biological) networks usually
has to be studied by means of Monte Carlo (MC) algorithms to solve the master
equation, often called kinetic MC methods~\cite{kinMC0,kinMC,Fichthorn91}.
Still, the
example given here is illustrative for the more complex PPTIS method
advocated in this paper.  The PPTIS method  combines the iterative
solution
of Eq.~(\ref{eq:recurs2}) for the overall rate constant with the
TIS algorithm~\cite{ErpMoBol2003}.  This approach will enable treatment of
a much wider variety of systems with a diffusive character, but not
with such a rigid  structure as the one dimensional Markov chain.

\subsection{TIS formalism}\label{TIStheory}

Let $x_t$ denote a point in phase space defined by the position $r$
and momenta $p$ of all  particles in the system at time $t$,
$x_t \equiv \{r(t),p(t) \}$.  Although the expressions derived in this
paper are also valid for stochastic dynamics, we assume here that the
system is deterministic: $x_t=f(x_{t^\prime},t-t^\prime)=f(x_0,t)$.
Evaluation of the time propagator function $f(x,t)$ requires the
integration of motion (e.g. by means of MD) starting from
configuration $x$ over a time interval $t$.

The TIS method is based on the measurement of the flux though dividing
surfaces. For this purpose we define a set of $n$ non-intersecting
multidimensional interfaces $\{0,1 \dots n \}$ described by an order
parameter $\lambda(x)$ which is a function of the phase space point
$x$.
In this way, interface $i$ is the ${\mathcal N}-1$ dimensional surface
$\{x|\lambda(x)=\lambda_i \}$ for a system with ${\mathcal N}$ degrees of freedom.
We choose $\lambda_i$, $i=0\dots n$ such that $\lambda_{i-1} < \lambda_i$, 
and that the
boundaries of state A and B are described by $\lambda_0$ and
$\lambda_n$, respectively.  To derive the TIS rate expression we need
to introduce characteristic functions that do not only depend on the
instantaneous position, but on the whole trajectory $x_t$.  For each
phase point $x$ and each interface $i$, we define a backward time
$t_i^b(x)$ and forward time $t_i^f(x)$:
\begin{eqnarray}
t_i^b(x_0) &\equiv& -\max \left[ \{ t | \lambda(x_t) = \lambda_i
\wedge t \leq 0 \} \right] \nonumber \\ t_i^f(x_0) &\equiv& +\min
\left[ \{ t | \lambda(x_t) = \lambda_i \wedge t \ge 0 \} \right],
\end{eqnarray}
which mark the points of first crossing
with interface $i$ on a forward (backward) trajectory starting in
$x_0$.  Note that $t_i^b$ and $t_i^f$ defined in this way always have
positive values. Following Ref.~[\onlinecite{ErpMoBol2003}], we then introduce
two-fold characteristic functions that depend on two interfaces
$i \neq \nobreak j$.
\begin{eqnarray}
\bar{h}_{i,j}^b(x) & = &
\begin{cases}
1 \quad \textrm{ if } t_i^b(x) < t_j^b(x) ,\\ 0 \quad \textrm{
otherwise}
\end{cases} \nonumber \\
\bar{h}_{i,j}^f(x) & = &
\begin{cases}
1 \quad \textrm{ if } t_i^f(x) < t_j^f(x) ,\\ 0 \quad \textrm{
otherwise}
\end{cases} \nonumber \\
\end{eqnarray}
which measure whether the backward (forward) time evolution of $x$
will reach interface $i$ before $j$ or not.
However, as the interfaces do not intersect, the time evolution has to
be evaluated only for those phase points $x$ that are in between the
two interfaces $i$ and $j$. In case $i<j$, we know in advance that
$t_i^{b,f}(x) < t_j^{b,f}(x)$ if $\lambda(x)<\lambda_i$ and
$t_i^{b,f}(x) > t_j^{b,f}(x)$ if $\lambda(x)>\lambda_j$.  When the
system is ergodic, both interfaces $i$ and $j$ will be crossed in
finite time and thus $\bar{h}_{i,j}^b(x) + \bar{h}_{j,i}^b(x) =
\bar{h}_{i,j}^f(x) + \bar{h}_{j,i}^f(x)=1$.  The two backward
characteristic functions define the TIS {\it overall} states
${\mathcal A}$ and ${\mathcal B}$:
\begin{eqnarray}\label{TISstates}
h_{\mathcal A}(x)=\bar{h}_{0,n}^b(x), \quad h_{\mathcal
B}(x)=\bar{h}_{n,0}^b(x).
\end{eqnarray}
Together, the {\it overall} states cover the entire phase space and,
within certain limits, do not sensitively depend on the precise
boundaries of stable states $A$ and $B$ as long as they are
reasonable.  That is, the stable states $A$ and $B$ should not
overlap, each path from $A$ to $B$ should be a true reaction for the
case of interest, and the chance that a trajectory starting in $A$ and
ending in $B$ will return to $A$ should be as unlikely as a complete
new event.  Within this formalism the rate constant can be written as
\cite{ErpMoBol2003}:

\begin{equation}
k_{AB}= \frac{\left \langle h_{\mathcal A}( x_0 ) \dot{h}_{\mathcal
B}( x_0 ) \right \rangle }{ \left \langle h_{\mathcal A}(x_0) \right
\rangle},
\label{rateTIS}
\end{equation}
where the dot denote the time derivative at $t=0$ and the brackets
$\left \langle \ldots \right \rangle$ denote the equilibrium ensemble
averages.  In the above equation, one can replace $x_0$ by $x_t$ for
any $t$ and we will often skip the argument $x_0$ when it does not
lead to
confusion. Eq.~(\ref{rateTIS}) measures the effective flux through the
phase space
hyper-surface dividing ${\mathcal A}$ from ${\mathcal B}$.  To express this
effective flux  in terms that can be
computed  we  define the general flux function
\begin{eqnarray} \label{fluxij}
\phi_{ij}(x_0) = \bar{h}_{j,i}^b(x_0) \lim_{\Delta t \rightarrow 0}
\frac{1}{\Delta t} \theta\big( \Delta t- t_i^f(x_0) \big)
\end{eqnarray}
with $\theta(x)$ the Heaviside step-function.
In principle, the flux expressions are defined in the limit $\Delta t
\rightarrow 0$ where it converges to: $\phi_{ij}=\bar{h}_{j,i}^b
|\dot{\lambda}| \delta( \lambda(x)-\lambda_i )$.
In practice, however,
Eq.~(\ref{fluxij}) will be more convenient with $\Delta t$ equal to
the MD time step.  This function measures the velocity through
interface $i$ at $t=0$ while coming directly from $j$ without having
recrossed $i$.

With this notation we can write Eq.~(\ref{rateTIS}) as
\begin{equation}
\label{eq:kab_phi}
k_{AB}= \left \langle \phi_{n,0} \right \rangle / \left \langle
h_{\mathcal A} \right \rangle = \left \langle \phi_{i,0}
\bar{h}^f_{n,0} \right \rangle / \left \langle h_{\mathcal A} \right
\rangle 
\end{equation}
for each $ 0 \leq i \leq n$, The ensemble average $\left< \phi_{i,0}
\right> $ is the effective positive flux from $A$ through $i$.  This
means that we only count
those phase points that will cross interface $i$ in the positive
direction in one $\Delta t$ time step, and come directly from $A$,
or equivalently, $x$ should be a first crossing point of the
corresponding trajectory that starts in $A$.

For the reverse reaction rate, we can write similar expressions, but
then related to the effective negative flux:
\begin{equation}
\label{eq:kba_phi}
k_{BA}= \left \langle \phi_{0,n} \right \rangle / \left \langle
h_{\mathcal B} \right \rangle = \left \langle \phi_{i,n}
\bar{h}^f_{0,n} \right \rangle / \left \langle h_{\mathcal B} \right
\rangle.
\end{equation}
Note that this effective flux formalism has a lot of flexibility.  The
second equality in Eq.~(\ref{eq:kab_phi}) and (\ref{eq:kba_phi}) is true
for any interface $\lambda_i$, independently on its position on the
barrier or shape.  If transition state theory (TST) is valid and we
could choose
our order parameter function $\lambda(x)$ such
that $\{x|\lambda(x)=\lambda_i\}$ is exactly the transition state dividing
surface, all points on this surface directing to $B$
would contribute to the rate.
In that case, Eq.~(\ref{eq:kab_phi}) would become
equivalent to the TST expression~\cite{DC78,Dellago02}.  However, for
complex systems it is almost impossible to determine the
multidimensional dividing surface or it would require a lot of a-priori
knowledge.  Instead, the strategy of TIS is to relate the effective
positive flux through one interface to that through one closer to $A$.

By introducing the weighted ensemble average $\left \langle g(x)
\right \rangle_\omega$ for an observable $g(x)$ and a weight function
$\omega(x)$:
\begin{eqnarray}
\left \langle g(x) \right \rangle_{\omega} \equiv \frac{\left \langle
g(x) \omega(x) \right \rangle}{\left \langle \omega(x) \right \rangle}
\end{eqnarray}
we can rewrite the rate expression~(\ref{eq:kab_phi}) into a product of
terms that can be measured as conditional ensemble averages.
\begin{eqnarray}
\label{rate4}
k_{AB} & = & \frac{ \left \langle \phi_{1,0} \right \rangle}{ \left
\langle h_{\mathcal{A}}\right \rangle} \left\langle \bar{h}_{n,0}^f
\right \rangle_{\phi_{1,0}} \nonumber \\ &=& \frac{ \left \langle
\phi_{1,0} \right \rangle}{ \left \langle h_{\mathcal{A}}\right
\rangle} \prod_{i=1}^{n-1} \left\langle \bar{h}_{i+1,0}^f \right
\rangle_{\phi_{i,0}}.
\end{eqnarray}
In the last equality we have used the TIS relation
(Eq.~(16) in Ref.~[\onlinecite{ErpMoBol2003}]).  The strength of the rate
expression (\ref{rate4}) is that it
rewrites Eq.~(\ref{rateTIS}) as a product of conditional probabilities
each factor being much higher than the final rate.
This recast expression  allows a better
accessible route for the computational approach, as it
drastically reduces the number of necessary MC moves required
for an accurate calculation of $k_{AB}$.
The actual algorithm consists of the computation of the effective
flux from $A$ through $\lambda_1$, $\left \langle \phi_{1,0} \right
\rangle/ \left \langle h_{\mathcal{A}}\right \rangle$, by means of
standard MD, followed by the determination of the conditional
probabilities $ \langle \bar{h}_{i+1, 0}^f \rangle_{\phi_{i,0}}$ via a
path sampling procedure. The advantage of TIS over the TPS rate
constant algorithm is that the path length is variable so that each
path can be limited to its strict necessary minimum length.  Moreover,
the effective positive flux formalism ensures that only positive terms
are accounted in the Monte Carlo scheme, which improves the
convergence.  Still, for diffusive systems path lengths can become
exceedingly long, making an accurate calculation problematic, even
when using TIS.

\subsection{PPTIS formalism}\label{PPTIStheory}
In previous section, we have reformulated the TIS theory in a way that
facilitates the step toward PPTIS.  It is important to note that the
PPTIS formalism is also based on a relation between effective fluxes,
however, not only in the positive, but also in the negative direction.
The algorithm allows a more efficient evaluation of the forward rate
$k_{AB}$ and, besides, also gives the reverse rate $k_{BA}$ with
negligible extra costs.

\begin{figure}[b]
\includegraphics[width=6cm, angle=-90]{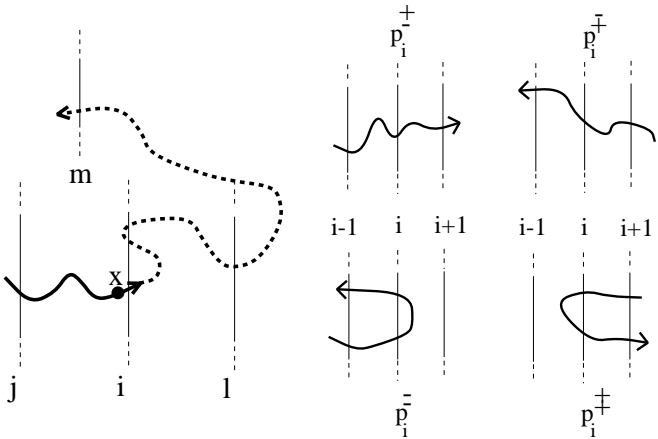}
\caption{Visualization of the different conditional crossing
probabilities $P(_m^l|_j^i)$ and $(p_i^\pm,p_i^=,
p_i^\mp,p_i^\ddagger)$.  Left: Explanation for $P(_m^l|_j^i)$. The
condition on $x$, $|_j^i)$, is represented with the solid line: From
$x$ we must cross interface $\lambda_i$ in one $\Delta t$ time step
and, besides, when propagating backward in time it should not cross
$\lambda_i$ another time before crossing $\lambda_j$. A possible
continuation of the trajectory corresponding to a positive
contribution to $(_m^l|$ is given by the dashed line.  The evaluation
of $P(_m^l|_j^i)$ requires the evaluation of all possible $x$
fulfilling the condition and the measurement of the fraction for which the
forward propagation crosses $\lambda_l$ before $\lambda_m$. This
probability can be evaluated for any possible set of four interfaces
$\{\lambda_i, \lambda_j, \lambda_l$, $\lambda_m \}$.  However, if we
keep the interfaces $\lambda_i, \lambda_j$ and $\lambda_l$ on its
position and place $\lambda_m$ between $\lambda_i$ and $\lambda_l$ we
get the trivial case $P(_m^l|_j^i)=0$. Similarly, we get
$P(_m^l|_j^i)=1$ if we place $\lambda_m$ at the right side of
$\lambda_l$. Right: visualization of the one-interface crossing
probabilities $(p_i^\pm,p_i^=, p_i^\mp,p_i^\ddagger)$.  Possible
trajectories that correspond to a positive contribution of these
probabilities are shown.}
\label{figprobvisu}
\end{figure}

It is convenient to introduce a short notation for the effective flux
function
\begin{eqnarray}
\Phi_{ij}^{lm} (x) \equiv \phi_{ij}(x) \bar{h}_{l,m}^f(x)
\end{eqnarray}
In this notation, the ensemble average of $\Phi_{i,0}^{n,0}$ is the
effective positive flux from $A$ through $\lambda_i$ going to
$B$. Renormalizing with $\langle \phi_{ij}\rangle$ defines the
conditional probabilities
  
\begin{eqnarray}
P(_m^l|_j^i) \equiv {\left< \Phi_{ij}^{lm} \right>}/ {\left< \phi_{ij}
\right>}.
\end{eqnarray}
In words, this is the probability for the system to reach interface
$l$ before $m$ under the condition that it crosses at $t=0$ interface
$i$, while coming directly from interface $j$ in the past (see
Fig.~\ref{figprobvisu}).  The rate constants can now be written in
terms of these probabilities
\begin{eqnarray}\label{ratePPTIS}
k_{AB}=\frac{ \left \langle \phi_{1,0} \right \rangle}{ \left \langle
h_{\mathcal{A}}\right \rangle} P(_0^n|_0^1), \quad k_{BA}=\frac{ \left
\langle \phi_{n-1,n} \right \rangle}{ \left \langle
h_{\mathcal{B}}\right \rangle} P(_{n}^0|_{n}^{n-1}), \quad
\end{eqnarray}
In addition, we define the one-interface crossing probabilities
$p_i^\pm,p_i^\mp,p_i^=,$ and $p_i^\ddagger$.
\begin{eqnarray}
p_i^\pm & \equiv & P(_{i-1}^{i+1}|_{i-1}^i) , \qquad p_i^\mp \equiv
P(_{i+1}^{i-1}|_{i+1}^i) \nonumber \\ p_i^= & \equiv &
P(_{i+1}^{i-1}|_{i-1}^i), \qquad p_i^\ddagger \equiv
P(_{i-1}^{i+1}|_{i+1}^i),
\label{onehop}
\end{eqnarray}
which fulfill the following relation:
\begin{eqnarray}
p_i^\pm+p_i^= = p_i^\mp+p_i^\ddagger=1,
\label{pi_unit}
\end{eqnarray}
A schematic visualization of $P(_m^l|_j^i)$ and the probabilities
$(p_i^\pm,p_i^=, p_i^\mp,p_i^\ddagger)$ is given in
Fig.~\ref{figprobvisu}.
We also define long-distance crossing probabilities $P_i^+$ and
$P_i^-$, similar to those in Sec.~\ref{illusex}
\begin{eqnarray}
P_i^+ \equiv P(_0^i|_0^1), \qquad P_i^- \equiv P(_{i}^0|_{i}^{i-1}).
\label{longhop}
\end{eqnarray}

The main assumption in PPTIS is that trajectories lose their memory,
over a short time, and hence over a short ``distance'', as measured by
\ $\lambda$.  We require that the interfaces are set such that no
memory effects are present over more than the distance between two
interfaces or, equivalently, that following relation is obeyed:
\begin{eqnarray}
\left \langle g(x) \right \rangle_{\phi_{i,i\pm q}} \approx \left
\langle g(x) \right \rangle_{\phi_{i,i\pm 1}},
\label{memorycond}
\end{eqnarray}
with $q$ an integer larger than one and $g(x)$ any observable
corresponding to the actual state $x$ or any future state.  With this
assumption we can derive recursive relations for the long-distance
crossing probabilities using the PPTIS concept introduced in
Sec.~\ref{illusex} (see Appendix \ref{AppendixB}):
\begin{eqnarray}
P_j^+ &=& \frac{p_{j-1}^\pm P_{j-1}^+}{p_{j-1}^\pm+p_{j-1}^=
P_{j-1}^-} \nonumber \\ P_j^- &=& \frac{p_{j-1}^\mp
P_{j-1}^-}{p_{j-1}^\pm+p_{j-1}^= P_{j-1}^-}
\label{centralrec}
\end{eqnarray}
To solve these recursive expressions we start with $P_1^+ = P_1^-=1$,
after which we iteratively determine $(P_j^+,P_j^-)$ for $j=2,\ldots$
until $j=n$.  Substitution of the long distance crossing probabilities
into Eq.~(\ref{ratePPTIS}) results in
\begin{eqnarray}\label{ratePPTIS2}
k_{AB}=\frac{ \left \langle \phi_{1,0} \right \rangle}{ \left \langle
h_{\mathcal{A}}\right \rangle} P_n^+, \quad k_{BA}=\frac{ \left
\langle \phi_{n-1,n} \right \rangle}{ \left \langle
h_{\mathcal{B}}\right \rangle} P_{n}^-, \quad
\end{eqnarray}

We obtain the reverse rate and the equilibrium constant $C=k_{AB}/k_{BA}$ without any significant
extra costs, whereas in other path sampling methods the calculation of
the reverse rate would require another comparable computational
effort~\cite{Dellago02}.

\subsection{The Sampling}
\label{secalgorithm}
The PPTIS method requires the determination of the
$p_i^\pm,p_i^=,p_i^\mp$, and $p_i^\ddagger$ probabilities. However,
$p_i^\pm$ and $p_i^=$ are defined in a different ensemble than
$p_i^\mp$ and $p_i^\ddagger$.  In most cases, it will be convenient to
calculate the four probabilities simultaneously.  Therefore, we define an ensemble that includes both
ensembles via the weight function $\phi_{i\pm}(x)$:
\begin{eqnarray}
\phi_{i\pm}(x) \equiv \phi_{i,i-1}(x)+\phi_{i,i+1}(x)
\end{eqnarray}

In this ensemble, $p_i^\pm$ and $p_i^\mp$ equal
\begin{eqnarray}
p_i^\pm=\frac{\left \langle \Phi_{i,i-1}^{i+1,i-1} \right
\rangle_{\phi_{i\pm}}}{ \left \langle \phi_{i,i-1} \right
\rangle_{\phi_{i\pm}}}, \qquad p_i^\mp=\frac{\left \langle
\Phi_{i,i+1}^{i-1,i+1} \right \rangle_{\phi_{i\pm}}}{ \left \langle
\phi_{i,i+1} \right \rangle_{\phi_{i\pm}}},
\end{eqnarray}
and $p_i^=$ and $p_i^\ddagger$ follow from Eq.~(\ref{pi_unit}).

For a correct sampling of phase points $x_0$ in this ensemble, we
generate all possible paths starting from interface ${i-1}$ or ${i+1}$
and ending either by crossing ${i-1}$ or ${i+1}$ with at least one
crossing with $i$. The sampling is performed using the shooting move
as explained in Ref.~\cite{ErpMoBol2003} with the difference that there is no
need to reject the backward integration as it is allowed to reach
either ${i-1}$ or ${i+1}$. All paths are confined within
$\lambda_{i-1}$ and $\lambda_{i+1}$ and have, even in the case of
multiple $\lambda_{i}$ crossings, only one time-slice $x$ along the
path for which $\phi_{i\pm}(x) \neq 0$. This defines the phase point $x_0$.  A big
advantage is that the time reversal moves become now very efficient,
as they are cheap and will always result in a new phase point $x_0$ of
the ensemble.

\subsection{Position of the Interfaces}
\label{secposint}
Contrary to the TIS technique, where the interfaces should be close to
obtain good statistics, the interfaces should be sufficient apart in
the PPTIS method to ensure complete loss of memory.  A simple test for
Eq.~(\ref{memorycond}) would be to measure
$\left \langle g(x) \right \rangle_{\phi_{i,i-1}}$ for different
separations between $\lambda_i$ and $\lambda_{i-1}$. The velocity
$\dot{\lambda}$ at the crossing point through $\lambda_{i}$ would be a
good candidate for the function $g$~\footnote{As not only the average
velocity should be the same $\left \langle \dot{\lambda}(x) \right
\rangle_{\phi_{i,i-q}}=\left \langle \dot{\lambda}(x) \right
\rangle_{\phi_{i,i-1}}$, but the whole distribution of velocities at
$\lambda_i$, we used in Sec.~\ref{secNum} the velocity distribution
overlap as measure of the memory loss.}.  This test is time consuming
if it has to be applied for  all possible interface separations.
However, one can estimate the memory loss for interface separations smaller
than the chosen one during the rate constant calculation. If
the interfaces are sufficient apart, one 
obtains a reasonable validation that the memory is vanished before reaching
the next surface.
Substituting $\dot{\lambda}(x) $ into
Eq.~(\ref{memorycond}) gives

\begin{eqnarray}
\label{criterion2}
\left \langle \dot{\lambda}(x_0) \right \rangle_{\phi_{i+1,i}} \approx
\left \langle \dot{\lambda}(x_0)) \right \rangle_{\phi_{i+1,i-1}}
\end{eqnarray}
This relation can be rewritten in the ensemble of $\phi_{i\pm}$:
\begin{eqnarray}
\frac{\left \langle \dot{\lambda}(x_F) \bar{h}_{{i+1},{i-1}}^f (x_0)
\right \rangle_{\phi_{i\pm}} }{\left \langle \bar{h}_{{i+1},{i-1}}^f
(x_0) \right \rangle_{\phi_{i\pm}} } \approx \frac{\left \langle
\dot{\lambda}(x_F) \Phi_{i,i-1}^{i+1,i-1} (x_0) \right \rangle_{\phi_{i\pm}}
}{ \left \langle \Phi_{i,i-1}^{i+1,i-1}  (x_0)\right \rangle_{\phi_{i\pm}} }
\label{critdir+}
\end{eqnarray}
where $x_F \equiv f\big(x_0,\min[t_{{i-1}}^f(x_0),t_{{i+1}}^f(x_0)] \,
\big)$ is the paths endpoint and $\dot{\lambda}(x_F)$ its velocity.  A
similar expression can be derived for the reverse direction.
The endpoint velocity $\dot{\lambda}(x_F)$ is indicatory for the path's likelihood to
progress along the order parameter $\lambda$. Therefore, we can reasonably expect
that if Eq. (\ref{critdir+}) is true for all interfaces $\lambda_i$,
the systematic error in the overall crossing probability $P_n^+$ due to the memory loss assumption will be small.
Criterion (\ref{critdir+}) is obeyed if the endpoint velocities of the
$(-+)$- and $(++)$-paths are the same, or if there are no $(++)$-paths present
at all. The first case is true if the barrier is relatively flat and the
interfaces are sufficiently far apart. The second case is typical for the
system  going uphill. If the system is going downhill without having reached
the basin of attraction of the product state, the memory loss requirement will
demand a careful examination on both the order parameter and the interface
positions.

\begin{figure}[t]
\includegraphics[width=6cm]{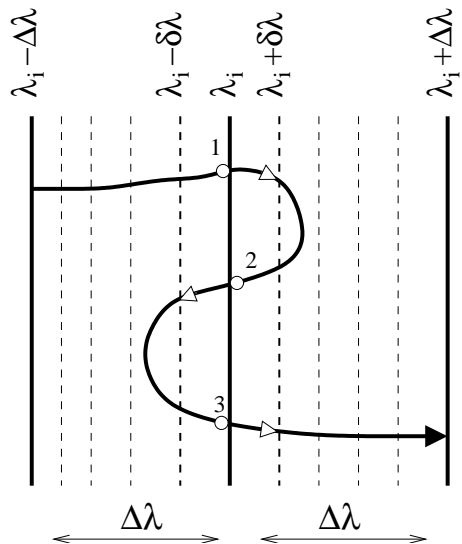}
\caption{Illustration for the calculation of the MLF on the grid of
sub-interfaces. One possible path is shown confined between
$\lambda_{i-1}$ and $\lambda_{i+1}$. }
\label{subint}
\end{figure}

An quantitative
indication of the fulfillment of the memory loss criterion can be
obtained by defining a memory-loss-function (MLF), for instance the
ratio of the two terms at both sides of the equality in
Eq.~(\ref{critdir+}). 
If we use a fine grid of $n_{sub}$ 
sub-interfaces between $\lambda_{i-1}$ and $\lambda_{i+1}$ (See
Fig.~\ref{subint}), we   can  measure this function with 
a resolution of $\delta \lambda = \Delta \lambda / 
n_{sub}$ with $\Delta \lambda \equiv 
\lambda_i-\lambda_{i-1}=\lambda_{i+1}-\lambda_{i}$~\footnote{To simplify 
notation we assume here an equidistant interface separation for all
interfaces. One is, however, by no means restricted to do so and one can place
each interface at an optimum position concerning efficiency, memory loss and
ergodic sampling.}.
The function
$MLF_i(j \delta \lambda)$ with $j=1 \ldots n_{sub}$ can be calculated  in the 
$\phi_{i\pm}$ ensemble during a PPTIS simulation.
To do this we will need an additional MC move
when the path has multiple recrossings with  interface $i$.
On 
the path in Fig.~\ref{subint} 
between $\lambda_{i-1}$ and $\lambda_{i+1}$ only one phase
point (1) belongs to the ensemble $\phi_{\lambda_i \pm \Delta \lambda} \equiv
\phi_{i\pm}$. However, in the ensemble defined by the two most inner
sub-interfaces $\phi_{\lambda_i\pm \delta \lambda}$ three points belong to the
ensemble (1,2 and 3). Therefore, an additional MC move is required to
measure MLF$(j \delta \lambda)$ for $j \delta \lambda< \Delta \lambda$.
This works as
follows: For each path in the $\phi_{i\pm}$ ensemble, we  loop over all
sub-interfaces $j$. For each $j$, first collect all the phase
points that belong to the ensemble of $\phi_{\lambda_i \pm j \delta \lambda}$. 
Secondly,  sample the MLF function for a random  point 
out of the $n$ points $\{x_0^{(1)}, x_0^{(2)}, \ldots, x_0^{(n)} \}$ for
which $\phi_{\lambda_i \pm j \delta \lambda}(x_0) \neq 0$. 
Third, take a uniform random number $\alpha$ between $[0:1]$ and repeat the second step if $\alpha > 1/n$. Otherwise,
continue the loop over $j$ until $j=n_{sub}$. Finally,  generate a new path, and repeat the whole procedure.
 
\subsection{Comparing TIS with PPTIS} \label{comparison}

In order to make a proper efficiency comparison between the 
two methods, we need to
estimate the computational effort for a certain fixed error.
We rather calculate the error in the equilibrium constant
$C=k_{AB}/k_{BA}$ instead of in the rate $k_{AB}$ itself because  
the expression of $C$ in terms of the averages, that have to be calculated 
separately, is much simpler than the recursive expression~(\ref{centralrec})
of $k_{AB}$.
Hence, the error propagation from the error in the individual terms  
is simpler and yields a more transparent comparison with TIS.
As $P^+_j/P^-_j=P^+_{j-1}/P^-_{j-1} \cdot p^\pm_{j-1}/p^\mp_{j-1}$, the
equilibrium constant $C$ can be written as:
\begin{eqnarray}\label{CPPTIS}
C_{\rm PPTIS} &=& \frac{[ \langle\phi_{1,0}/\langle h_{\mathcal{A}}\rangle]}
{[\langle\phi_{n-1,n}\rangle/\langle h_{\mathcal{B}}\rangle]}
\Big[\frac{p^\pm_{n-1}}{p^\mp_{n-1}}\Big] \cdots \Big[
\frac{p^\pm_{1}}{p^\mp_{1}}\Big]
\end{eqnarray}

Each term within brackets $[\ldots]$ is calculated separately together
with its error. The error propagation of the total $n+1$ terms
determines the final overall error. Similarly, in TIS the expression
for $C$ can be written as:
\begin{eqnarray}\label{CTIS}
C_{\rm TIS} = \frac{[ \langle\phi_{1,0}/\langle h_{\mathcal{A}}\rangle]}
{[\langle\phi_{n-1,n}\rangle/\langle h_{\mathcal{B}}\rangle]}
\frac{[\mathcal{P}_A (\lambda_n | \lambda_{n-1})] \cdots
[\mathcal{P}_A (\lambda_2 | \lambda_1)]} {[\mathcal{P}_B (\lambda_0 |
\lambda_1)] \cdots [\mathcal{P}_B (\lambda_{n-2} | \lambda_{n-1})]}
\end{eqnarray}
with $\mathcal{P}_{A} (\lambda_j | \lambda_i) \equiv P(_0^j|_0^i)$ and
$\mathcal{P}_{B} (\lambda_j | \lambda_i) \equiv P(_n^j|_n^i)$.  Here,
in total $2n$ simulations have to be performed, each on a different
ensemble.  In practice, however, not all the interface ensembles are
needed, as $\mathcal{P}_A (\lambda_i | \lambda_{i-1})$ and
$\mathcal{P}_B (\lambda_i | \lambda_{i+1})$ will converge to unity in
the limit $i \rightarrow n$ and $i \rightarrow 0$,
respectively. Interestingly, for PPTIS the error in the terms
$[p^\pm_{i}/p^\mp_{i}]$ will be more or less the same for all $i$ on
the barrier. For TIS, however, the error in $[ \mathcal{P}_{A/B}
(\lambda_i | \lambda_{i-1})] $ will decrease when 
its value 
gets closer to
unity. In contrast, the path length required for the calculation of
these crossing probabilities will increase, while, again, being more
or less constant in PPTIS.  In Sec.~\ref{effscaling}, we will show
that the final TIS computation time, that involves these two effects,
scales quadratically as function of its barrier length and only
linearly for PPTIS.

\section{Numerical Results}
\label{secNum}

\subsection{The model system}
In Refs.~[\onlinecite{TPS99}] and [\onlinecite{ErpMoBol2003}], 
the TPS and TIS methods were tested on a bistable diatomic molecule immersed in a  fluid of purely repulsive
particles. Here, we use a very similar  system to test the PPTIS method,
consisting of $N$ two-dimensional particles interacting via the
Weeks-Chandler-Andersen (WCA) potential~\cite{WCA71}
\begin{equation}
V_{WCA}(r)=
\begin{cases}
4\epsilon [(r/\sigma)^{-12}-(r/\sigma)^{-6}] + \epsilon & \mbox{if }
\, r\leq r_0 \\ 0 \quad \quad & \mbox{if } r>r_0, \\
\end{cases}
\end{equation}
where $r$ is the interatomic distance, and $r_0\equiv
2^{1/6}\sigma$. In the following we will use reduced units so that the
energy and length parameters $\epsilon$ and $\sigma$, the mass of the
particles and the unit of time $(m\sigma^2/\epsilon)^{1/2}$ are all
equal to unity.  In addition, two of the $N$ particles are interacting
through a  double well potential
\begin{equation}\label{Vdiff}
V_{diff}(r)=
\begin{cases}
V_{dw}(r) & \mbox{if } r<r_{0}+w \\ 
h & \mbox{if } r_{0}+w<r<r_0+w+b \\ 
V_{dw}(r-b) & \mbox{if } r>r_0+w+b
\end{cases},
\end{equation}
where
\begin{equation}\label{Vdw}
V_{dw}(r)=h[1-(r-r_0-w)^2/w^2]^2.
\end{equation}
This  potential and its first derivative are continuous and the forces
are therefore well defined.  It has two minima at $r=r_0$, the compact
state or state $A$, and at $r=r_0+2w+b$, the extended state or state
$B$. The minima are separated by a total barrier of length $b+2w$ and
height $h$. For sufficiently large values of $h$, transitions between
the states become rare and the rate constants are well defined. For
sufficiently large values of $b$, trajectories on the barrier plateau
become diffusive.  We therefore expect this system to be a good test
case for the new PPTIS method.

We simulate the system at constant energy $E=1.0$ in a square box with
periodic boundary conditions. The number density is fixed at 0.7, by
adjusting the size of the box. The barrier length should always be
less than half the box's edge, implying the number of particles $N$
to increase accordingly with the value of the barrier length $b$. The
remaining barrier parameters are set to $h=15$ and $w=0.5$.  The total
linear momentum is conserved and is set to zero. The equations of
motion are integrated using the velocity Verlet algorithm with a time
step $\Delta t=0.002$.
The Monte Carlo path sampling is carried out both in PPTIS and TIS by
means of the shooting move and the path-reversal move, as explained in
Sec.~\ref{secalgorithm} and Ref.~[\onlinecite{ErpMoBol2003}]. 
The two moves were
performed with an equal probability of 50\%.  The momentum
displacement for the shooting move was always gaged such that the
acceptance ratio is about 40\%, which provides an optimum efficiency
of the sampling \cite{TPS99}.  The intermolecular distance $r$ is a
suitable order parameter to define the interfaces.

In the following two subsections we consider a system with a barrier
short enough to gather good statistics in a reasonable computation
time.  In section~\ref{effscaling} we study 
the gain in efficiency of PPTIS over TIS as a function of the
diffusive barrier length. In the final section~\ref{memloss} we test the memory loss assumption as
explained in Sec.~\ref{secposint}.

\subsection{System with short barrier}\label{shortbarr}

\begin{table}[b]
\begin{center}
\begin{tabular}{l*{3}{r@{$\pm$}l}}
& \multicolumn{2}{c}{$k_{AB}/10^{-10}$} &
\multicolumn{2}{c}{$k_{BA}/10^{-10}$} & \multicolumn{2}{c}{$C$} \\
\hline PPTIS & 2.75&0.07 & 1.95&0.04 & 1.41&0.05 \\ TIS & 2.8&0.2 &
2.03&0.06 & 1.4&0.1 \\
\end{tabular}
\end{center}
\caption{Comparison of PPTIS and TIS. Forward and backward rate
constants as well as the equilibrium constant are reported for the
system with short energy barrier.}
\label{restable}
\end{table}
We simulated a system of $N=100$ WCA particles with a barrier length
$b=2$. The minima of $V_{diff}(r)$ are located at $r\simeq 1.12$ and
$r\simeq 4.12$, and the diffusive plateau extends from $r\simeq 1.62$ to
$r\simeq 3.62$. 
State $A$ is defined by interface $\lambda_0$
as $r<1.22$ and state $B$ by interface $\lambda_{17}$ as $r>4.02$.
In the intermediate regime 16 interfaces were chosen 
at $r=1.24,$ 1.34, 1.40,1.46, 
1.52, 1.62, 2.02, 2.42, 2.82, 3.22, 3.62, 3.72, 3.78, 3.84, 3.90,  
and $4.00$. 

%

\begin{figure}[t]
\begin{center}
\includegraphics[width=8cm,keepaspectratio]{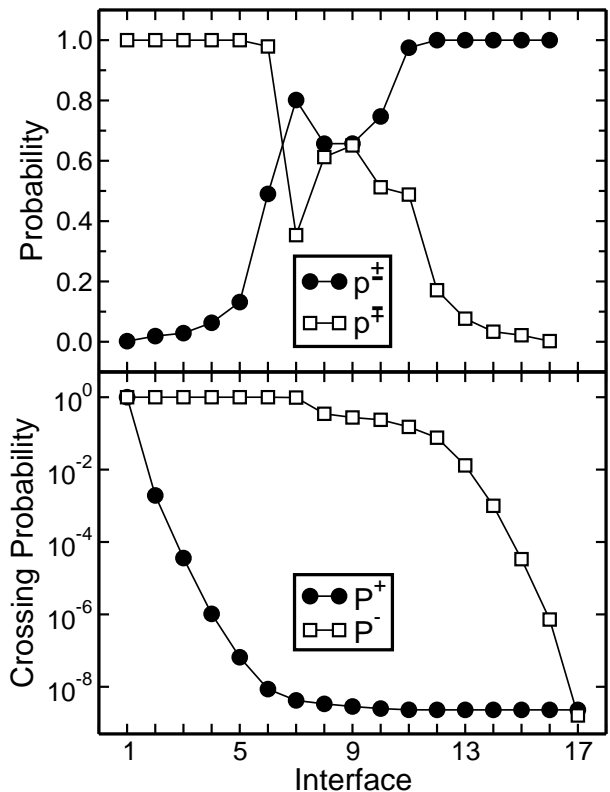}
\end{center}
\caption{
Top: PPTIS one-interface crossing probabilities $p^\pm$,
$p^\mp$, see Eq.~(\ref{onehop}). The $p^=$, $p^\ddag$ probabilities
follow directly from Eq.~(\ref{pi_unit}). Bottom:
PPTIS long-distance crossing probabilities $P_i^+$, $P_i^-$, see
Eq.~(\ref{longhop}). The last points contribute to the rate constants
as in Eq.~(\ref{ratePPTIS2}). In both graphs the error is within
symbol size.}
\label{PPTISprob}
\end{figure}

First, we ran straightforward MD simulations in state $A$ and $B$ to compute the
fluxes that appear in both Eq.~(\ref{CPPTIS}) and (\ref{CTIS})
by
counting the number of positive  crossings through interfaces
$\lambda_1$ and $\lambda_{16}$, respectively \cite{ErpMoBol2003}. 
We obtained the values $\langle\phi_{1,0}\rangle/\langle
h_{\mathcal{A}}\rangle=0.1160\pm 0.0008$ and
$\langle\phi_{16,17}\rangle/\langle h_{\mathcal{B}}\rangle=0.117\pm
0.001$.  
Subsequently, we calculated the conditional probabilities in
Eq.~(\ref{CPPTIS}). For PPTIS we 
calculated the one-interface crossing probabilities for all the 16 interfaces
on the barrier, 
while TIS simulations
show convergence after 11 windows 
for both the forward and the backward 
reaction path. 
In Fig.~\ref{PPTISprob} we report
the one-interface crossing probabilities $p^\pm_i$, $p^\mp_i$ 
and the long-distance crossing probability $P_i^+$, $P_i^-$.
The long-distance crossing probabilities appearing in the  rate constant 
Eq.~(\ref{ratePPTIS2}) 
for $n=17$ 
are $P^+_n=(2.37\pm0.06)10^{-9}$ and $P^-_n=(1.67\pm0.03)10^{-9}$.
These values can be compared with their TIS counterparts
$\mathcal{P}_A(\lambda_n|\lambda_1)=(2.4\pm0.2)10^{-9}$ and
$\mathcal{P}_B(\lambda_0|\lambda_{n-1})=(1.74\pm0.05)10^{-9}$.  We note
that because for the first 5 interfaces $i=1\ldots 5$, $p^\mp_i$ equals
unity, $P^-_i$ is constant up to $i=6$.  Similarly, for $i=11\ldots
16$, $p^\pm_i$ is unity and $P^+_i$ shows a plateau starting at
$i=12$. This means that in the PPTIS methods, although for the
equilibrium constant $C$ all the windows are necessary, the separate
computation of $k_{AB}$ and $k_{BA}$ requires fewer windows. The
result is consistent with what we found in TIS.  We report in table
\ref{restable} the final rate and equilibrium constants.  They all
coincide within the statistical error.

\begin{figure}[t]
\begin{center}
\includegraphics[width=8cm,keepaspectratio]{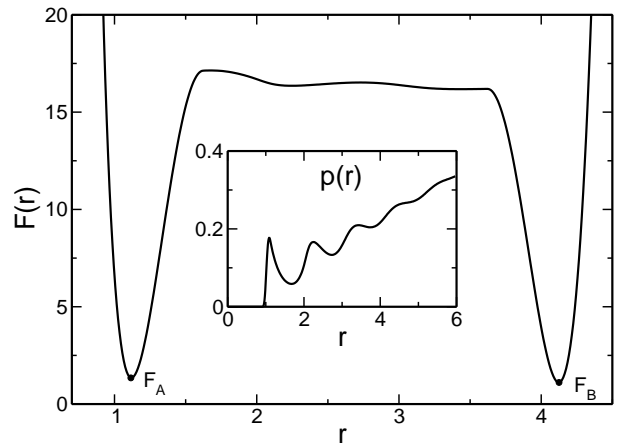}
\end{center}
\caption{The free energy as function of the dimer interatomic distance
$r$.  It was calculated from the dimer potential $V_{diff}(r)$,
Eq.~(\ref{Vdiff}), corrected by a factor involving $p(r)$, the
probability of finding the dimer atoms at distance $r$, as reported in
Eq.~(\ref{energyeq}). The function $p(r)$ was computed using a biased
MD simulation and is plotted in the inset.  From the minima of $F(r)$
we derived the free energy difference $\Delta F = F_A-F_B$ and hence
the equilibrium constant.}
\label{freeenergy}
\end{figure}

Another way 
to derive the equilibrium constant 
is by using
the relation $C=\exp(\Delta F/k T)$ where
$k$  is  Boltzmann's constant, $T$ is the temperature, and $\Delta F =
F_A-F_B$ is the free energy difference between states $A$ and $B$.  To
check our results 
we 
calculated the free energy as
function of the intermolecular distance $r$ using a straightforward MD
simulation with a bias potential given by $-V_{diff}(r)$. This is
equivalent to simulating a system of pure WCA particles and computing
$p(r)$, the probability of finding two particles at a distance $r$.  The
advantage is that in such a system it does not make any difference
which two molecules we consider to compute $p(r)$ and we can thus
increase the statistics by averaging on all pairs.  It can be
easily seen that
\begin{equation}\label{energyeq}
F(r)=V_{diff}(r)-k T \ln p(r) + constant
\end{equation}
where in our microcanonical simulations the temperature $T$ is
derived from average kinetic energy.
A plot of the free energy is shown in Fig.~\ref{freeenergy}. From this
curve we derive the free energy difference $\Delta F$ between the two
minima, and the equilibrium constant $C=1.369\pm 0.001$.
All results are consistent with each other within
the statistical error.

\begin{figure}[t]
\begin{center}
\includegraphics[width=8cm,keepaspectratio]{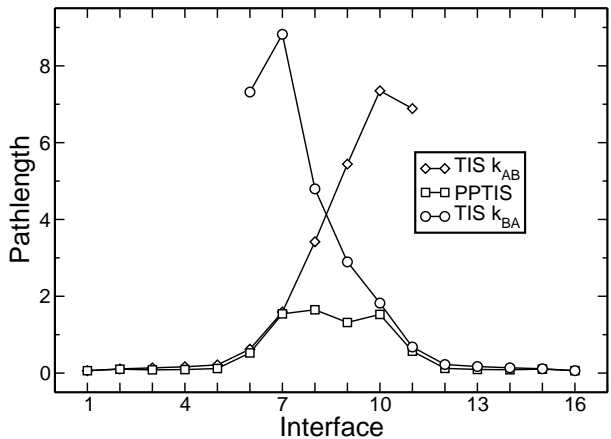}
\end{center}
\caption{Comparison of path-lengths for PPTIS and the TIS simulations
for the calculation of the forward and backward rate constant. Because
of the diffusive character of the system, the TIS path-lengths keep
growing as the interface moves further from the initial stable
state. The PPTIS path-lengths on the contrary stay constant. The
errors are within the symbol size.
}
\label{pathlength}
\end{figure}

\subsection{Efficiency Scaling }\label{effscaling}

In both the PPTIS and the TIS method the final equilibrium constant is
a product of factors  given  by Eqs.~(\ref{CPPTIS}) and (\ref{CTIS}), respectively. We determined each factor independently by  performing $M$ simulation blocks of
$m$ Monte Carlo cycles. We adjusted $m$ so that the relative standard
deviation of each term in Eqs.~(\ref{CPPTIS}) and (\ref{CTIS}) after
$M$ block averages was an arbitrary value of 3\%. We measured, under
the same computational conditions (1.4 GHz AMD Athlon), the CPU-time
required and summed up all the times to get the relative
efficiency. The final errors on the rate constants given above were obtained by
standard propagation rules using {\em all} the available blocks of
simulations.

We computed the  computation times to reach the prefixed 3\% error  for
each factor in Eqs.~(\ref{CPPTIS}) and (\ref{CTIS}) and found  that for the simple dimer 
system the efficiency of PPTIS is at least a factor two higher than TIS.
 In figure \ref{pathlength}
we plot the average path-length in each window
for the two
methods. The direct comparison shows that on the barrier PPTIS keeps
the path length constant while the TIS path length increases. This is
expected but it does not directly imply a gain in efficiency 
as
the relative error in the TIS terms 
is  smaller for the longer paths.


\begin{figure}[b]
\begin{center}
\includegraphics[width=8cm,keepaspectratio]{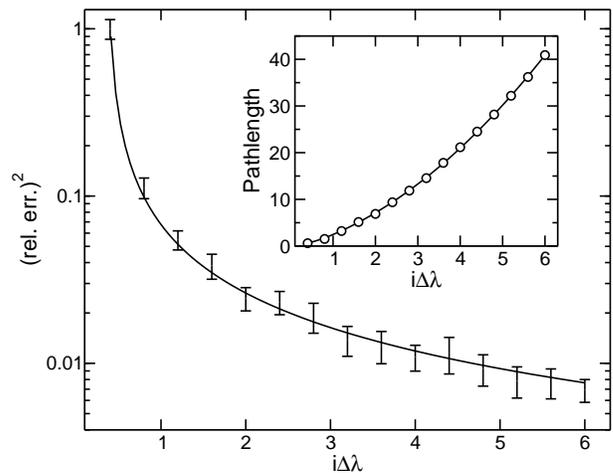}
\end{center}
\caption{
The relative variance for the TIS crossing probability  $\mathcal{P}_A(\lambda_{i+1}|\lambda_i)$ plotted  as function of the barrier length for a
system with total barrier length $b=6$. The relative variances have
been rescaled to the one of the first 
interface measured from the start of the plateau. 
We fitted the relative variance  with an inverse linear function.
Inset: the
the average path length for these simulations as function of the barrier length.
The error bar is within symbol size.  The solid line
is a  second-order polynomial fit.}
\label{effi}
\end{figure}
In order to compare the effciency of both  PPTIS and   TIS quantitatively  as a function of barrier length 
we consider the windows $i=1\ldots N_W$. In each window we perform simulations of $m$ Monte Carlo cycles. Let the average path length in each window be $L_i$ and 
relative error in  the observable (e.g. hopping probability) be $\epsilon_i$. If we assume that $m$ is large enough so that all simulations are uncorrelated, 
then $\epsilon_i$ scales as the inverse
square root of $m$.  To obtain a fixed error $\epsilon$, one has to rescale
the number of paths by $(\epsilon_i/\epsilon)^2$. 
Moreover, the acceptance ratio is almost independent 
of the path length for TIS in the kind of systems we have studied here
\footnote{This is contrast with what was found in Ref.~[\onlinecite{bolhuis03}] 
using TPS. There, we found a dramatic  decrease of acceptance as function of 
the barrier length. This is caused by increasing rejection probability
due to the constraint that the path should not be larger than the fixed length.
TIS is less sensitive to this as the path length is variable. Still, for more
complex free energy landscapes that may contain many entropic barriers,
the TIS acceptance probability might also decrease when the path lengths
get longer.}.
As a result we found that the required CPU time for $m$ 
MC cylces scales
linearly with its average path length.
The total computation time is then proportional to
\begin{equation}
m \sum_{i=1}^{N_W} L_i \left(\frac{\epsilon_i}{\epsilon}\right)^2
\end{equation}
If the barrier is very long we can neglect the initial and final
windows 
on the steep side of the barrier 
and consider only those on the plateau
for which we assume 
a fixed interface separation $\Delta \lambda=b/N_W$.
The PPTIS method keeps $L_i$ and $\epsilon_i$ 
more or less
constant 
(see Fig.~\ref{pathlength} ).
The efficiency $\eta$ is the ratio of the TIS to the PPTIS computation
time
\begin{equation}\label{effieq}
\eta
\equiv \frac{\textrm{CPU}_{\rm TIS}}{\textrm{CPU}_{\rm PPTIS}}
\propto
\frac{
\sum_{i=1}^{N_W} L_i \epsilon_i^2
}{N_W}
\end{equation}
where $L_i$ and $\epsilon_i$ are now the TIS path-length and relative
error for window $i$.  To study the behavior of $L_i$ and $\epsilon_i$
we focus on the TIS calculations for the forward reaction rate
constant $k_{AB}$.  The observables are the probabilities
$\mathcal{P}_A(\lambda_{i+1}|\lambda_i)$ (see Eq.~(\ref{CTIS})). 

To estimate the TIS effective computation time as function of the barrier
length, we first examined the model of Sec.~\ref{illusex}
with $t_{i,i\pm1}=\frac{1}{2}$ (uniform symmetric hopping). 
This simplified system allows us to obtain some analytical results and to
perform path sampling on wide barriers with millions of paths. 
We found that the relative error $\epsilon_j$ in the long distance hopping
probabilities $T[1\rightarrow_0^j]$ scales as $\sim \frac{1}{\sqrt j}$.
Moreover, the average length of the corresponding path 
(the number of hops) scales quadratically with $j$, while the acceptance ratio
remained constant.

To test  
whether this
scaling behavior 
 also applies to the dimer model,
we considered a system of
$N=256$ particles with a barrier length $b=6$. The minima of
$V_{diff}(r)$ are located at $r\simeq 1.12$ and $r\simeq 8.12$.  State
$A$ is defined by interface $\lambda_0$ as $r<1.20$. We defined 22
other interfaces, 16 of which on the barrier plateau from $r\simeq
1.62$ to $r\simeq 7.62$ at intervals
$\Delta \lambda =0.4$.
By running several TIS simulations, we computed the 
crossing probabilities $\mathcal{P}_A(\lambda_{i+1}|\lambda_i)$ and  their
standard deviations for
$i=1\ldots 21$.  In
Fig.~\ref{effi} we show the relative variance and the average path length for the windows on the barrier. Indeed, the  scaling behavior is as expected.
Evaluating the sum in
Eq.~(\ref{effieq}) yields that the relative efficiency $\eta$
scales linearly with the barrier width.
This means that while the  computation time scales quadratically with the barrier width for TIS it scales 
linearly for PPTIS.


%

\begin{figure}[b]
\begin{center}
\includegraphics[width=8cm,keepaspectratio]{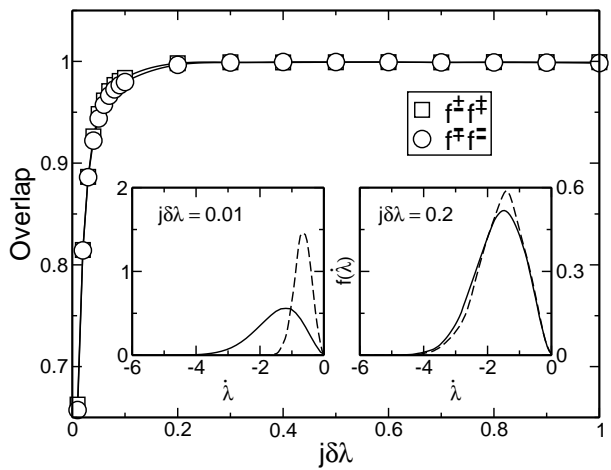}
\end{center}
\caption{Memory loss function computed using the overlap of the
distributions of the endpoint velocity $\dot{\lambda}$, see
Sec. \ref{memloss}. In the insets we plot the distributions for paths
of the $(+-)$ ensemble (solid line) and the $(- -)$ one (dashed line), for
two different window sizes
$j \delta \lambda =0.01$ and
$j \delta \lambda =0.2$. The first two distributions are different, and
the second ones are almost overlapping.
}
\label{overlap}
\end{figure}
\subsection{Validity of the memory loss assumption}\label{memloss}

We computed the memory loss function MLF($j \delta\lambda$) as defined
in section \ref{secposint} for the same $b=2$ system of section
\ref{shortbarr} using the method described in Sec.~\ref{secposint} with a
central interface at $r=2.62$ and $\delta\lambda=0.01$ and $j$ ranging from 1
to  $100$, corresponding to the entire length of the barrier plateau. Since
not only the mean
value of the endpoint velocity $\dot{\lambda}$ but its complete
probability distribution $f(\dot{\lambda})$ should be equal for paths
of the ensemble $(-+)$ and $(++)$ we computed the overlap
\begin{equation}
\int_{-\infty}^{+\infty}
\sqrt{f^\pm(\dot{\lambda})f^\ddag(\dot{\lambda})} \;d\dot{\lambda}.
\end{equation}
Similar expression was used for the paths $(+-)$ and $(--)$. The
results are reported in Fig.~\ref{overlap}. It can be seen that for
$j \delta\lambda\geq 0.2$ the memory loss assumption is satisfied.

In umbrella sampling methods, where the phase space is divided into
partially overlapping windows which are later matched, the choice of
the windows is a trade-off between the diffusion of the order parameter
and its equilibration time in the window~\cite{FrenkelSmit}.  Because
of the first effect the size of the window should be chosen as small
as possible, while the second puts a lower bound on it. Similar
considerations apply to PPTIS with the addition of the memory loss requirement,
which is also a lower bound to the window size.  Taking
all this into accounts we believe $\Delta\lambda=0.2$ is an optimal
choice. 
 The optimal value of $\Delta\lambda$ is of course dependent on the  system and on the choice of order parameter, and has to be estimated by trial and error.

\section{Discussion and conclusions}
\label{secConclusions}

In this paper we have introduced a path sampling algorithm for the
efficient calculation of rate constants of two state activated
processes with a diffusive barrier.  The method is based on the
division of phase space by interfaces. We then calculate
hopping probabilities from one interface to another, using transition
path sampling shooting moves and time reversal moves 
as our basic instrument to create new
paths on each interface ensemble. Using either the iterative scheme
given here or for more general hopping networks 
the method of kinetic Monte Carlo, one can solve the
master equation and obtain the final forward and backward rate
constants.  In deriving this algorithm we assumed complete memory loss
between interface, such that the system becomes essentially Markovian,
thus validating the use of kinetic Monte Carlo and similar algorithms.
We showed that for a relatively simple system, the diatomic molecule,
the memory loss assumption (loss of correlation) holds over the entire
barrier. We expect that for more complex systems this memory loss requirement 
will certainly be fulfilled, provided that 
the dynamics has a stochastic character and
the interfaces are placed sufficiently 
far apart. 
However, the choice of  order parameter requires still some caution, 
possibly more than in TIS, in order to satisfy the memory loss requirement.
For the  simple dimer  system, we showed that  PPTIS is 
already twice as  fast as  TIS.  More importantly, we argued that the 
computation time
scales linearly with the barrier length, instead of quadratically 
as for TIS and maybe even with a higher power for TPS.  
This opens up possibilities for accurate rate constant 
calculations for complex activated processes.

The method advocated here to tackle  diffusive barriers in complex systems 
is not the first one that has been proposed
in the literature.  Several techniques have been put forward in the
last decade, for instance the diffusive barrier algorithm by 
Ruiz-Montero et al.~\cite{Ruiz} and the coarse MD method by 
Hummer and Kevrekidis\cite{hummer}.
The latter technique uses short trajectories
to calculate the average force projected on a order parameter
space. They use that force to integrate a stochastic equation of
motion and explore the free energy landscape in that way. 
Rate constants can then in principle  be obtained from the dynamics on 
this coarse  grained surface. 

 The method of  Ruiz-Montero et al. is in essence a reactive
flux method but  enhances the statistics by measuring the flux on
many different places on the barrier and weigh those contributions
such that the error in the rate constant is as small as possible. 
The weighing function
turns out to be proportional to the inverse of the 
barrier free energy profile. This
means that to get a meaningful result one should have access to the
free energy landscape on the barrier, before the rate constant
calculation. However, due to complexity, the order parameters chosen
as reaction coordinate are not necessarily correct, sometimes
resulting in inaccurate barriers and very small transmission
coefficients. 
Moreover, the calculation of a transmission coefficient suffers
from the same quadratically dependence of the barrier length.

We stress that there is a large difference between the reactive
flux method based on transition state theory and the  PPTIS technique. 
Although we use
hyper-surfaces to divide the phase-space we do not rely on a global
large transmission coefficient. Instead, we calculate local transmission
coefficients and use those as hopping probabilities.
We believe that the PPTIS method can be applied to sample diffusive pathways 
and calculate rate constants in many different complex systems behaving more 
or less diffusive, such as protein folding, nucleation, chemical  reactions, 
biochemical networks, and gas diffusion through membranes. 
In the near future we intend to improve possible sampling problems occurring  
when  the used order parameter is a very bad reaction coordinate.

\section{Appendix A}
\label{AppendixA}

In this appendix we will derive the recursive relations (\ref{metahop}) for the
chain of metastable states. For the transfer in the positive direction
we can write
\begin{eqnarray}
T[1 \rightarrow_0^j] &=&T[1 \rightarrow_0^{j-1}] T[j-1
\rightarrow_0^{j}]\nonumber \\ &=&T[1 \rightarrow_0^{j-1}] \big(
1-T[j-1 \rightarrow_j^{0}] \big)
\label{A1}
\end{eqnarray}

and for the reverse direction

\begin{eqnarray}
T[j-1 \rightarrow_{j}^0] &=& t_{j-1,j-2} T[j-2\rightarrow_{j}^0]
\nonumber \\ &=& t_{j-1,j-2}\Big( T[j-2\rightarrow_{j-1}^0] +
\nonumber \\ & &T[j-2\rightarrow_0^{j-1}] T[j-1\rightarrow_{j}^0]
\Big) \nonumber \\
\label{A2}
&=& t_{j-1,j-2} \Big(T[j-2\rightarrow_{j-1}^0] + \\ & &\big(1-T[j-2
\rightarrow_{j-1}^0] \big) T[j-1\rightarrow_{j}^0] \Big) \nonumber
\end{eqnarray}

Bringing the $T[j-1\rightarrow_{j}^0]$ terms of Eq.~(\ref{A2}) to the
left side gives us:
\begin{eqnarray}
T[j-1\rightarrow_{j}^0] =\frac{t_{j-1,j-2} T[j-2\rightarrow_{j-1}^0]
}{ 1-t_{j-1,j-2}(1- T[j-2\rightarrow_{j-1}^0]) }
\label{A3}
\end{eqnarray}
Using $1-t_{j-1,j-2}=t_{j-1,j}$, we see that Eq.~(\ref{A3}) is
equivalent to the second line in Eq.~(\ref{metahop}).  The first line
of Eq.~(\ref{metahop}) is then obtained by the substitution 
into Eq.~(\ref{A1}).

\section{Appendix B}
\label{AppendixB}
The criterion of Eq.~(\ref{memorycond}) gives for any positive integer
$q > 0$ following approximate relations:
\begin{eqnarray}
P(_{m}^{l}|_{i \pm q}^i) \approx P(_{m}^{l}|_{i \pm 1}^i) \nonumber \\
P(_{i- 1}^{i+ q}|_{i + 1}^i) \approx P(_{i - 1}^{i+ q}|_{i - 1}^i)
(p_i^\ddagger/p_i^\pm) \nonumber \\ P(_{i+ 1}^{i- q}|_{i - 1}^i)
\approx P(_{i + 1}^{i- q}|_{i + 1}^i) (p_i^=/p_i^\mp)
\end{eqnarray}

With this in mind we can start a derivation similar to Appendix
\ref{AppendixA}:
\begin{eqnarray}
P_j^+ & \equiv & P(_0^j|_0^1)= P(_0^{j-1}|_0^1)P(_0^j|_0^{j-1})
\nonumber \\ & \approx & P(_0^{j-1}|_0^1)P(_0^j|_{j-2}^{j-1})
\nonumber \\ &= & P_{j-1}^+ \Big( 1- P(_j^0|_{j-2}^{j-1}) \Big)
\nonumber \\ & \approx & P_{j-1}^+ \Big( 1- P(_j^0|_{j}^{j-1})
\frac{p_{j-1}^=}{p_{j-1}^\mp} \Big) \nonumber \\ & = & P_{j-1}^+ \Big(
1- P_{j}^- \frac{p_{j-1}^=}{p_{j-1}^\mp} \Big)
\label{AppenB1}
\end{eqnarray}

and for the reverse direction we can write:

\begin{eqnarray}
P_j^- & = & P(_{j}^0|_{j}^{j-1})= p_{j-1}^\mp P(_{j}^0|_{j}^{j-2})
\nonumber \\ & \approx & p_{j-1}^\mp P(_{j}^0|_{j-1}^{j-2}) \nonumber
\\ & \approx & p_{j-1}^\mp \big[P(_{j-1}^0|_{j-1}^{j-2})+
P(_{0}^{j-1}|_{j-1}^{j-2}) P(_{j}^0|_{j-2}^{j-1}) \big] \nonumber \\ &
= & p_{j-1}^\mp \big[ P_{j-1}^-+ \big(1-P_{j-1}^- \big)
P(_{j}^0|_{j-2}^{j-1}) \big] \nonumber \\ & \approx & p_{j-1}^\mp
\big[ P_{j-1}^-+ \big(1-P_{j-1}^- \big) P(_{j}^0|_{j}^{j-1})
\frac{p_{j-1}^=}{p_{j-1}^\mp} \big] \nonumber \\ & = & p_{j-1}^\mp
\big[ P_{j-1}^-+ \big(1-P_{j-1}^- \big) P_j^-
\frac{p_{j-1}^=}{p_{j-1}^\mp} \big]
\end{eqnarray}

Bringing the $P_j^-$ terms to the left results in:
\begin{eqnarray}
P_j^- = \frac{ P_{j-1}^- p_{j-1}^\mp }{ 1- \big(1- P_{j-1}^- \big)
p_{j-1}^= }
\end{eqnarray}
With the help of Eq.~(\ref{pi_unit}) we can see that this is
equivalent to expression (\ref{centralrec}). Substitution of this
relation into Eq.~(\ref{AppenB1}) results in the expression for
$P_j^+$ in Eq.~(\ref{centralrec}).


\bibliographystyle{prsty}
\bibliography{paper}

\end{document}